\documentclass[11pt]{article}
\usepackage{amssymb}
\usepackage{amsmath}
\usepackage[dvips]{graphicx}
\usepackage[usenames]{color}

\newtheorem{theorem}{Theorem}
\newtheorem{corollary}{Corollary}
\newtheorem{proposition}{Proposition}
\newtheorem{lemma}{Lemma}
\newtheorem{example}{Example}
\newtheorem{definition}{Definition}
\newtheorem{remark}{Remark}
\newcommand{\beq}{\begin{equation}}
\newcommand{\eeq}{\end{equation}}
\newcommand{\beas}{\begin{eqnarray*}}
\newcommand{\eeas}{\end{eqnarray*}}
\newcommand{\bea}{\begin{eqnarray}}
\newcommand{\eea}{\end{eqnarray}}
\newcommand{\bei}{\begin{itemize}}
\newcommand{\eei}{\end{itemize}}
\newcommand{\ben}{\begin{enumerate}}
\newcommand{\een}{\end{enumerate}}
\newcommand{\bet}{\begin{theorem}}
\newcommand{\eet}{\end{theorem}}
\newcommand{\bel}{\begin{lemma}}
\newcommand{\eel}{\end{lemma}}
\newcommand{\bep}{\begin{proposition}}
\newcommand{\eep}{\end{proposition}}
\newcommand{\bed}{\begin{definition}}
\newcommand{\eed}{\end{definition}}
\newcommand{\bec}{\begin{corollary}}
\newcommand{\eec}{\end{corollary}}
\newcommand{\bex}{\begin{example}}
\newcommand{\eex}{\end{example}}

\newenvironment{proof}[1][Proof]{\noindent\textbf{#1.} }{\
  \rule{0.5em}{0.5em}}
\begin{document}
\title{$L_1$ penalized LAD estimator for high dimensional linear regression}
\author{Lie Wang\thanks{Department of Mathematics, Massachusetts Institute of
Technology, Cambridge, MA 02139, USA; e-mail: {\tt
liewang@math.mit.edu} Research supported by  NSF Grant
DMS-1005539.}}
\date{}
\maketitle
\begin{abstract}
In this paper, the high-dimensional sparse linear regression model is considered,
where the overall number of variables is larger than the number of observations. We investigate the $L_1$ penalized least absolute deviation method. Different from most of other methods, the $L_1$ penalized LAD method does not need any knowledge of standard deviation of the noises or any moment assumptions of the noises. Our analysis shows that the method achieves near oracle performance, i.e. with large probability, the $L_2$ norm of the estimation error is of order $O(\sqrt{k\log p/n})$. The result is true for a wide range of noise distributions, even for the Cauchy distribution.  Numerical results are also presented.

\end{abstract}
\newpage

\section{Introduction}
High dimensional linear regression model, where the number of observations is much less than the number of unknown coefficients, has attracted much recent interests in a number of fields such as applied math, electronic engineering, and statistics. In this paper, we consider the following classical high dimensional linear model:
\begin{equation}
\label{model} Y=X\beta+z.
\end{equation}
where $Y=(y_1,y_2,\cdots,y_n)'$ is the $n$ dimensional vector of outcomes, $X$ is the $n\times p$ design matrix, and $z=(z_1,z_2,\cdots,z_n)'$ is the $n$ dimensional vector of measurement errors (or noises). We assume $X=(X_1,X_2,\cdots,X_p)$ where $X_i\in R^n$ denotes the $i$th regressor or variable. Throughout, we assume that each vector $X_i$ is normalized such that $\|X_i\|_2^2=n$ for $i=1,2,\cdots,p$. We will focus on the high dimensional case where $p\geq n$ and our goal
is to reconstruct the unknown vector $\beta\in R^p$.

Since we are considering a high dimensional linear regression problem, a key assumption is the sparsity of the true coefficient $\beta$. Here we assume,
$$T=supp(\beta) \text{ has } k<n \text{ elements.} $$
The set $T$ of nonzero coefficients or significant variables is unknown. In what follows, the true parameter value $\beta$ and $p$ and $k$ are implicitly indexed by the
sample size $n$, but we omit the index in our notation whenever this does not cause confusion.

Ordinary least square method is not consistent in the setting of $p>n$. In recent years, many new methods have been proposed to solve the high dimensional linear regression problem. Methods based on $L_1$ penalization or constrained $L_1$ minimization have been extensively studied. Dantzig selector was proposed in \cite{CT2007}, which can be written as
$$\hat\beta_{DS}=\arg\min_{\gamma\in R^p} \|\gamma\|_1, \text{ subject to } \|X'(Y-X\gamma)\|_{\infty}\leq c\sigma\sqrt{2n\log p} ,$$
for some constant $c>1$. It is clear that the Dantzig selector depend on the standard deviation of the noises and the Gaussian assumption. General constrained $L_1$ minimization methods for noiseless case and Gaussian noise were studied in \cite{CWX2010a}. More results about the constrained $L_1$ minimization can be found in for example \cite{CRT2006}, \cite{D2006}, \cite{CWX2010b} and the references therein.

Besides the constrained minimization methods, the lasso ($L_1$ penalized least square) type methods have been studied in a number of papers, for example, \cite{T1996}, \cite{BRT2009}, and \cite{MY2009}. The classical lasso estimator can be written as
$$\hat\beta_{lasso}=\arg\min_{\gamma}\frac{1}{2}\|Y-X\gamma\|_2^2+\lambda\|\gamma\|_1,$$
where $\lambda$ is the penalty level (tuning parameter). In the setting of Gaussian noise and known variance, it is suggested in \cite{BRT2009} that the penalty could be $$\lambda=2c\sigma\sqrt{n\Phi^{-1}(1-\alpha/2p)},$$
where $c>1$ is a constant and $\alpha$ is small chosen probability. By using this penalty value, it was shown that the lasso estimator can achieve near oracle performance, i.e. $\|\hat\beta_{lasso}-\beta\|_2\leq C(k\log (2p/\alpha)/n)^{1/2}$ for some constant $C>0$ with probability at least $1-\alpha$.

The lasso method has nice properties, but it also replies heavily on the Gaussian assumption and a known variance. In practice, the Gaussian assumption may not hold and the estimation of the standard deviation $\sigma$ is not a trivial problem. In a recent paper, \cite{BCW2011} proposed the square-root lasso method, where the knowledge of the distribution or variance are not required. Instead, some moment assumptions of the errors and design matrix are needed. Other than the constrained optimization or penalized optimization methods, the stepwise algorithm are also studied, see for example \cite{Z2009} and \cite{CW2011}. It is worth noting that to properly apply the stepwise methods, we also need assumptions on the noise structure or standard deviation of the noises.

It is now seen that for most of the proposed methods, the noise structure plays an important role in the estimation of the unknown coefficients. In most of the existing literatures, either an assumption on the error distribution or a known variance is required. Unfortunately, in the high dimensional setup, these assumptions are not always true. Moreover, in cases where heavy-tailed errors or outliers are found in the response, the variance of the errors may be unbounded. Hence the above methods cannot be applied.

To deal with the cases where the error distribution is unknown or may has heavy tail. We propose the following $L_1$ penalized least absolute deviation ($L_1$ PLAD) estimator.
\begin{equation}
\label{lasso1} \hat\beta\in\arg\min\{\gamma:
\|Y-X\gamma\|_{1}+\lambda\|\gamma\|_1\}.
\end{equation}

The least absolute deviation (LAD) type of methods are important when heavy-tailed errors present. These methods have desired robust properties in linear regression models, see for example \cite{BK1978}, \cite{H1981} and \cite{PK1997}. Recently, the penalized version of the LAD method was studied. Variable selection properties and consistency of the $L_1$ penalized LAD were discussed in for example \cite{WLJ2007}, \cite{GH2010}, and \cite{LZ2011}.

In this paper, we present analysis for the $L_1$ PLAD method and we discuss the  selection of penalty level, which does not depend on any unknown parameters or the noise distribution. Our analysis shows that the $L_1$ PLAD method has surprisingly good properties. The main contribution of the present paper has twofold. (1) We proposed a rule for setting the penalty level, it is simply
$$\lambda=c\sqrt{2A(\alpha)n\log p},$$
where $c>1$ is a constant, $\alpha$ is a chosen small probability, and $A(\alpha)$ is a constant such that $2p^{-(A(\alpha)-1)}\leq \alpha$. In practice, we suggest to take $c=1.1$ or we can simply choose $\lambda=\sqrt{2n\log p}$, see the numerical study section for more discussions. This choice of penalty is universal and we only assume that the noises have median 0. (2) We show that with high probability, the estimator has near oracle performance, i.e. with high probability
$$\|\hat\beta-\beta\|_2=O(\sqrt{\frac{k\log p}{n}}).$$
It is important to notice that we do not have any assumptions on the distribution or moments of the noise. Actually, even for Cauchy distributed noise, where the first order moment does not exist, our results still hold.

Importantly, the problem retains global convexity, making the method computationally efficient.  Actually, we can use ordinary LAD method package to solve the $L_1$ penalized LAD estimator. This is because if we consider the penalty terms as new observations, i.e. $Y_{n+i}=0$ and $x_{n+i,j}=\lambda\times I(j=i)$ for $i,j=1,2,\cdots, p$. Then our $L_1$ penalized estimator can be considered as an ordinary LAD estimator with $p$ unknown coefficients and $p+n$ observations. Hence it can be solved efficiently.

The rest of the paper is organized as follows. Section 2 discusses the choice of penalty level. In section 3, the main results about the estimation error and several critical lemmas are presented. We also briefly explain the main idea of the proofs. Section 4 presents the simulation study results, which shows the $L_1$ penalized LAD method has very good numerical performance regardless the noise distribution. Technical lemmas and the proofs of theorems are given in section 5.

\section{Choice of Penalty}

In this section, we discuss the choice of the penalty level for the $L_1$ PLAD estimator. For any $\gamma\in R^p$, let $Q(\gamma)=\|Y-X\gamma\|_{1}$. Then the $L_1$ PLAD estimator can be written as
$$ \hat\beta\in\arg\min\{\gamma: Q(\gamma)+\lambda\|\gamma\|_1\}.$$
An important quantity to determine the penalty level is the sub-differential of $Q$ evaluated at the point of true coefficient $\beta$. Recall that the measurement errors $z_i$ follow some continuous distribution with median 0. Assume that $z_i\neq 0$ for all $i$, then the sub-differential of $Q(\gamma)=\|Y-X\gamma\|_1$ at point $\gamma=\beta$ can be written as
$$S=X'(sign(z_1),sign(z_2),\cdots,sign(z_n))',$$
where $sign(x)$ denotes the sign of $x$, i.e. $sign(x)=1$ if $x>0$, $sign(x)=-1$ if $x<0$, and $sign(0)=0$. Let $I=sign(z)$, then $I=(I_1,I_2,\cdots,I_n)'$ where $I_i=sign(z_i)$. Since $z_i$'s are independent and have median 0, we know that $P(I_i=1)=P(I_i=-1)=0.5$ and $I_i$ are independent.

The sub-differential of $Q(\gamma)$ at the point of $\beta$, $S=X'I$, summaries the estimation error in the setting of linear regression model.  We will choose a penalty $\lambda$ that dominates the estimation error with large probability. This principle of selecting the penalty $\lambda$ is motivated by \cite{BRT2009} and \cite{BCW2011}. It is worth noting that this is a general principle of choosing the penalty and can be applied to many other problems. To be more specific, we will choose a penalty $\lambda$ such that it is greater than the maximum absolute value of $S$ with high probability, i.e. we need to find a penalty level $\lambda$ such that
\begin{equation}
\label{penalty.con}
P(\lambda\geq c\|S\|_{\infty})\geq 1-\alpha,
\end{equation}
for a given constant $c>1$ and a given small probability $\alpha$. Note that $c$ is a theoretical constant and in practice we can simply take $c=1.1$. Since the distribution of $I$ is known, the distribution of $\|S\|_{\infty}$ is known for any given $X$ and does not depend on any unknown parameters.

Now for any random variable $W$ let $q_{\alpha}(W)$ denote the $1-\alpha$ quantile of $W$. Then in theory, $q_{\alpha}(\|S\|_{\infty})$ is known for any given $X$. Therefore if we choose $\lambda=cq_{\alpha}(\|S\|_{\infty})$, inequality (\ref{penalty.con}) is satisfied.

In practice, it might be hard to calculate the exact quantile $q_{\alpha}(\|S\|_{\infty})$  for a given $X$. One possible way to calculate or approximate it is by simulation, but this will cause additional computation time. Here we propose the following asymptotic choice of penalty.
\begin{equation}
\label{A.choice}
\lambda=c\sqrt{2A(\alpha)n\log p},
\end{equation}
where $A(\alpha)>0$ is a constant such that $2p^{-(A(\alpha)-1)}\leq\alpha$.

To show that the above choice of penalty satisfies (\ref{penalty.con}), we need to bound the tail probability of $\sum_{i=1}^{n}X_{ij}I_i$ for $i=1,2,\cdots,p$. This can be done by using the Hoeffding's inequality, see for example \cite{H1963}, and union bounds. We have the following lemma.
\begin{lemma}
\label{penalty.cho}
The choice of penalty $\lambda=c\sqrt{2A(\alpha)n\log p}$ as in (\ref{A.choice}) satisfies $$P(\lambda\geq c\|S\|_{\infty})\geq 1-\alpha.$$
\end{lemma}

From the proof previous lemma, we can see that if we use the following special choice of $\lambda$,
\begin{equation}
\label{A.choice2}
\lambda=2c\sqrt{n\log p},
\end{equation}
Then we have that
\begin{equation}
P(\lambda\geq c\|S\|_{\infty})\geq 1-\frac{2}{p}.
\end{equation}

The above penalties are simple and have good theoretical properties. Moreover, they do not require any conditions on matrix $X$ or value of $p$ and $n$. But in practice, since the bounds here are not very tight, these penalty levels tend to be relatively large and can cause additional bias to the estimator. It is worth pointing out that if there exists an $i\in \{1,2,\cdots,p\}$ such that $\|X_i\|_1<\lambda$, then $\hat\beta_i$ must be 0. Otherwise we can replace $\hat\beta_i$ by 0, and the value of $Q(\hat\beta)+\lambda\|\hat\beta\|_1$ will reduce by at least $(\lambda-\|X_i\|_1)|\hat\beta_i|$. This means if the penalty level $\lambda$ is too large, the $L_1$ PLAD method may kill some of the significant variables. To deal with this issue, we propose the following refined asymptotic choice of penalty level, provided some moment conditions on design matrix $X$.
\begin{lemma}
\label{A.choice3}
Suppose
\begin{equation}
B=\sup_{n}\sup_{1\leq j\leq p}\frac{1}{n}\|X_j\|_q^q<\infty,
\end{equation}
for some constant $q>2$. Assume $\Phi^{-1}(1-\alpha/2p)\leq (q-2)\sqrt{\log n}$. Then the choice of penalty $\lambda=c\sqrt{n}\Phi^{-1}(1-\frac{\alpha}{2p})$ satisfies $$P(\lambda\geq c\|S\|_{\infty})\geq 1-\alpha(1+\omega_n),$$
where $\omega_n$ goes to 0 as $n$ goes to infinity.
\end{lemma}

This choice of penalty replies on moment conditions of $X$ and relative size of $p$ and $n$, but it could be smaller than the previous ones and in practice it will cause less bias. We investigate the effect of different penalties in the numerical study section.

To simplify our arguments, in the following theoretical discussion we will use (\ref{A.choice2}) as the default choice of penalty.  It can be seen that the above choices of penalty levels do not depend on the distribution of measurement errors $z_i$ or unknown coefficient $\beta$. As long as $z_i$'s are independent continuous random variables with median 0, the choices satisfy our requirement. This is a big advantage over the traditional lasso method, which significantly relies on the Gaussian assumption and the variance of the errors.

\section{Properties of the Estimator}
In this section, we present the properties of the $L_1$ PLAD estimator. We shall state the upper bound for estimation error $h=\hat\beta-\beta$ under $L_2$ norm $\|h\|_2$. We shall also present the variable selection properties for both noisy and noiseless cases. The choice of penalty is described in the previous section. Throughout the discussion in this section, we assume the penalty $\lambda$ satisfies $\lambda\geq c\|S\|_{\infty}$ for some fixed constant $c>1$. In what follows, for any set $E\subset \{1,2,\cdots,p\}$ and vector $h\in R^p$, let $h_E=hI(E)$ denote the $p$ dimensional vector such that we only keep the coordinates of $h$ when their indexes are in $E$ and replace others by 0.

\subsection{Conditions on design matrix $X$}
We will first introduce some conditions on design matrix $X$. Recall that we assume $\lambda\geq c\|S\|_{\infty}$, this implies the following event, namely $h=\hat\beta-\beta$ belongs to the restricted set $\Delta_{\bar C}$, where
\begin{eqnarray*}
\Delta_{\bar C}=&&\{\delta\in R^p: \|\delta_{T}\|_1\geq \bar C \|\delta_{T^c}\|_1, \\
&&\text{ where } T\subset\{1,2,\cdots,p\} \text{ and } T \text{ contains at most } k \text{ elements}.\},
\end{eqnarray*}

and $\bar C=(c-1)/(c+1)$. To show this important property of the $L_1$ PLAD estimator, recall that $\hat\beta$ minimizes $\|X\gamma-Y\|_1+\lambda\|\gamma\|_1$. Hence
$$\|Xh+z\|_1+\lambda\|\hat\beta\|_1\leq \|z\|_1+\lambda\|\beta\|_1.$$
Let $T$ denote the set of significant coefficients. Then
\begin{equation}
\label{basic.1}
\|Xh+z\|_1-\|z\|_1\leq \lambda(\|h_{T}\|_1-\|h_{T^c}\|_1).
\end{equation}
Since the sub-differential of $Q(\gamma)$ at the point of $\beta$ is $X'I$, where $I=sign(z)$.
$$\|Xh+z\|_1-\|z\|_1\geq (Xh)'I\geq h'X'I\geq -\|h\|_1 \|X'I\|_{\infty}\geq -\frac{\lambda}{c}(\|h_{T}\|_1-\|h_{T^c}\|_1).$$
So
\begin{equation}
\|h_{T}\|_1\geq \bar C \|h_{T^c}\|_1,
\end{equation}
where $\bar C=\frac{c-1}{c+1}$.

The fact that $h\in \Delta_{\bar C}$ is extremely important for our arguments. This fact is also important for the arguments of classical lasso method and the square-root lasso method, see for example, \cite{BRT2009} and \cite{BCW2011}.

Now we shall define some important quantities of design matrix $X$. Let $\lambda_k^u$ be the smallest number such that for any $k$ sparse vector $d\in R^p$,
$$\|Xd\|^2_2\leq \lambda_k^u\|d\|^2_2.$$
Here $k$ sparse vector $d$ means that the vector $d$ has at most $k$ nonzero coordinates, or $\|d\|_0\leq k$. Similarly, let $\lambda_k^l$ be the largest number such that for any $k$ sparse vector $d\in R^p$,
$$\|Xd\|^2_2\geq \lambda_k^l\|d\|^2_2.$$
The definition of the above constants are essentially the Restricted Isometry Constants, see for example \cite{CT2005}, but we use different notations for upper and lower bounds. We also need to define the following restricted eigenvalues of design matrix $X$. These definitions are based on the idea of \cite{BRT2009}.
Let
\begin{eqnarray*}
&&\kappa_k^l(\bar C)=\min_{h\in \Delta_{\bar C}} \frac{\|Xh\|_1}{n\|h_T\|_2} \text{ and } \eta_k^l(\bar C)=\min_{h\in \Delta_{\bar C}} \frac{\|Xh\|_2}{\sqrt{n}\|h_T\|_2},\\
&&\kappa_k^u(\bar C)=\max_{h\in \Delta_{\bar C}} \frac{\|Xh\|_1}{n\|h_T\|_2} \text{ and } \eta_k^u(\bar C)=\max_{h\in \Delta_{\bar C}} \frac{\|Xh\|_2}{\sqrt{n}\|h_T\|_2}.
\end{eqnarray*}

To show the properties of the $L_1$ penalized LAD estimator, we need both $\kappa^l_k(\bar C)$ and $\eta^l_k(\bar C)$ to be bounded away from 0. To simplify the notations, when it is not causing any confusion, we will simply write $\kappa^l_k(\bar C)$ as $\kappa^l_k$, and $\eta^l_k(\bar C)$ as $\eta^l_k$.

\subsection{Important Lemmas}

Before presenting the main theorem, we first state a few critical lemmas.
From (\ref{basic.1}), we know that
$$\|Xh+z\|_1-\|z\|_1\leq \lambda\|h_{T}\|_1 .$$
To bound the estimation error, we shall first investigate the random variable $\frac{1}{\sqrt{n}}(\|Xh+z\|_1-\|z\|_1)$. For any vector $d\in R^p$, let
$$B(d)=\frac{1}{\sqrt{n}}\left|(\|Xd+z\|_1-\|z\|_1)-E(\|Xd+z\|_1-\|z\|_1) \right|.$$
We introduce the following important result.
\begin{lemma}
\label{Uniform.lemma}
Suppose $z_i$'s are independent random variables. Assume $p>n$ and $p>3\kappa_k^u$ then
\begin{equation}
P\left(\sup_{\|d\|_0=k, \|d\|_2=1} B(d)\geq (1+2C_1\sqrt{\lambda_k^u})\sqrt{2k\log p} \right)\leq 2p^{-4k(C_1^2-1)},
\end{equation}
where $C_1>1$ is a constant.
\end{lemma}

From the above lemma, we know that with probability at least $1-2p^{-4k(C_1^2-1)}$, for any $k$ sparse vector $d\in R^p$,
\begin{equation}
\frac{1}{\sqrt{n}}(\|Xd+z\|_1-\|z\|_1)\geq \frac{1}{\sqrt{n}}E(\|Xd+z\|_1-\|z\|_1)-C\sqrt{2k\log p}\|h\|_2,
\end{equation}
where $C=1+2C_1\sqrt{\lambda_k^u}$. This lemma shows that with high probability, the value of the random variable $\frac{1}{\sqrt{n}}(\|Xd+z\|_1-\|z\|_1)$ is very close to its expectation. Since the expectation is fixed and much easier to analysis than the random variable itself, this lemma plays an important role in our proof of the main theorem.

Next, we will investigate the properties of $E(\|Xd+z\|_1-\|z\|_1)$. We have the following lemmas.
\begin{lemma}
\label{l1.deriv}
For any continuous random variable $z_i$, we have that
$$\frac{d E(|z_i+x|-|z_i|)}{dx}=1-2P(z_i\leq -x).$$
\end{lemma}

Now we will introduce the scale assumptions on the measurement errors $z_i$. suppose there exists a constant $a>0$ such that
\begin{eqnarray}
&&P(z_i\geq x)\leq \frac{1}{2+ax} \text{ for all } x\geq 0\nonumber\\
&&P(z_i\leq x)\leq \frac{1}{2+a|x|} \text{ for all } x< 0 \label{scale.com}.
\end{eqnarray}
Here $a$ served as a scale parameter of the distribution of $z_i$. This is a very weak condition and even Cauchy distribution satisfies it. Based on this assumption, we have that for any $c>0$,
\begin{eqnarray*}
&&E(|z_i+c|-|z_i|)=c-2\int_{0}^{c}P(z_i<-x)dx\\
&\geq& c-2\int_{0}^{c}\frac{1}{2+ax}dx=c-\frac{2}{a}\log (1+\frac{a}{2}c).
\end{eqnarray*}
Hence we have the following lemma.
\begin{lemma}
\label{mean.l1.useful}
Suppose random variable $z$ satisfies condition (\ref{scale.com}), then
\begin{equation}
\label{Eofl1}
E(|z_i+c|-|z_i|)\geq \frac{a}{16}|c|(|c|\wedge \frac{6}{a}).
\end{equation}
\end{lemma}

\begin{remark}
This is just a weak bound and can be improved easily. But for simplicity, we use this one in our discussion.
\end{remark}

\subsection{Main Theorem}

Now we shall propose our main result. Here we assume that the measurement errors $z_i$ are independent and identically distributed random variables with median 0. We also assume that $z_i$s satisfy condition (\ref{scale.com}). Moreover, we assume $\eta_k^l>0$, $\kappa_k^l>0$ and
\begin{equation}
\label{Condition.I}
\frac{3\sqrt{n}}{16}\kappa_k^l> \lambda\sqrt{k/n}+C_1\sqrt{2k\log p}(1.25+\frac{1}{\bar C}),
\end{equation}
for some constant $C_1$ such that $C_1>1+2\sqrt{\lambda_k^u}$. We have the following theorem.
\begin{theorem}
\label{main}
Under the previous assumptions, the $L_1$ penalized LAD estimator $\hat\beta$ satisfies with probability at least $1-2p^{-4k(C_2^2-1)+1}$
\begin{equation*}
\|\hat\beta-\beta\|_2\leq  \sqrt{\frac{2k\log p}{n}}\frac{16(c\sqrt{2}+1.25 C_1+C_1/\bar C)}{a\eta_{k}^{l}}\sqrt{1+\frac{1}{\bar C}}.
\end{equation*}
where $C_1=1+2C_2\sqrt{\lambda^u_{k}}$ and $C_2>1$ is a constant.
\end{theorem}
\begin{remark}
From the proof of the theorem, we can see that the identically distributed assumption of the measurement errors is not essential. We just need that there exist a constant $a>0$ such that for all $i$, $P(z_i\geq x)\leq \frac{1}{2+ax}$ for  $x\geq 0$ and $P(z_i\leq x)\leq \frac{1}{2+a|x|}$ for $x< 0$. This is also verified in the section of simulation study.
\end{remark}

From the theorem we can easily see that asymptotically, with high probability,
\begin{equation}
\|\hat\beta-\beta\|_2=O(\sqrt{\frac{2k\log p}{n}}).
\end{equation}
This means that asymptotically, the $L_1$ PLAD estimator has near oracle performance and hence it matches the asymptotic performance of the lasso method with known variance.

A simple consequence of the main theorem is that the $L_1$ PLAD estimator will select most of the significant variables with high probability. We have the following theorem.
\begin{theorem}
Suppose $\hat T=supp(\hat\beta)$ be the estimated support of the coefficients. Then under the same conditions as in Theorem \ref{main}, with probability at least $1-2p^{-4k(C_2^2-1)+1}$,
\begin{equation}
\left\{i: |\beta_i|\geq \sqrt{\frac{2k\log p}{n}}\frac{16(c\sqrt{2}+1.25 C_1+C_1/\bar C)}{a\eta_{k}^{l}}\right\}\subset \hat T,
\end{equation}
where $C_1=1+2C_2\sqrt{\lambda^u_{k}}$ and $C_2>1$ is a constant.
\end{theorem}
\begin{remark}
This theorem shows that the $L_1$ PLAD method will select a model that contains all the variables with large coefficients. If in the main model, all the nonzero coefficients are large enough in terms of absolute value, then the $L_1$ PLAD method can select all of them into the model.
\end{remark}


A special but important case in high dimensional linear regression is the noiseless case. The next theorem shows that the $L_1$ PLAD estimator has nice variable selection property in the noiseless case.
\begin{theorem}
\label{thm.noiseless}
Consider the noiseless case. Suppose we use a penalty level $\lambda$ such that $\lambda<n\kappa_k^l(1)$, the $L_1$ penalized LAD estimator $\hat\beta$ satisfies $\hat\beta=\beta$.
\end{theorem}
\begin{remark}
Suppose $\kappa_k^l(1)$ are bounded away from 0 for all $n$ and we use the penalty level $\lambda=2\sqrt{n\log p}$. Then when $\sqrt{\log p}=o(n)$ and $n$ large enough. The $L_1$ penalized LAD estimator $\hat\beta$ satisfies $\hat\beta=\beta$.
\end{remark}

\section{Numerical Study}
In this section, we will show some numerical results. Throughout this section, we use $n=200$, $p=400$ and $k=5$ and set $\beta=(3,3,3,3,3,0,\cdots,0)$. We will study both the estimation properties and variable selection properties of the $L_1$ PLAD estimator under various noise structures. In our simulation study, we generate the design matrix $X$ by i.i.d. $N(0,1)$ random variables and then normalize the columns.

We first investigate the effect of different choices of penalty levels. Then we compare the $L_1$ PLAD method and the lasso method in the Gaussian noise case. We also study the numerical properties of $L_1$ PLAD estimator under different noise structures, including the heteroscedastic cases. We use the quantreg package and lars package in R to run the simulation.

\subsection{Effect of Penalty levels}
Section 2 discusses the choice of penalty levels. It is known that our desired choice is $cq_{\alpha}(\|S\|_{\infty})$. But since this value is hard to calculate, we propose several upper bounds and asymptotic choices. Now we will investigate the effect of different choices of penalty levels on the $L_1$ PLAD estimator. To be specific, we consider the following four penalties, $\lambda_1=\sqrt{1.5n\log p}$, $\lambda_2=\sqrt{2n\log p}$, $\lambda_3=\sqrt{3n\log p}$, and $\lambda_4=\sqrt{4n\log p}$. Note that they are all fixed choices and do not depend on any assumptions or parameters. For noises, we use (a) $N(0,1)$ noise, (b) $t(2)$ noise, and (c) Cauchy noise. For each setting, we run the simulation 200 times and the average $L_2$ norm square of the estimation errors are summarized in the following table.
\begin{table}[t]
\caption{The average of estimation error $\|\hat \beta-\beta\|_2^2$  over 200 simulations
under different penalty levels and error distributions. Numbers in the parentheses are the medians of the estimation errors of post $L_1$ PLAD method, i.e. results of ordinary LAD estimators on the selected subset.}
\label{penalty.levels}
\begin{tabular*}{\textwidth}{@{\extracolsep\fill}|c|c|c|c|c|}
\hline  & $\lambda_1$ & $\lambda_2$ & $\lambda_3$  & $\lambda_4$   \\
\hline $N(0,1)$ noise & 0.658 (0.356) & 1.054 (0.239) & 3.189 (0.095) & 23.730 (4.586) \\
\hline $t(2)$ noise  & 1.263 (0.552) & 2.351 (0.299) & 10.121 (0.081) & 33.018 (18.771)  \\
\hline  Cauchy noise  &  2.176 (0.861) & 4.736 (0.334) & 21.417 (0.103) & 39.351 (26.241)  \\
\hline
\end{tabular*}
\end{table}

From table \ref{penalty.levels} we can see that $\lambda_4$ is too large in our setup and it kills most of the variables. (It is worth noting that if we increase the sample size to for example $n=400$ and $p=800$, $\lambda_4$ becomes a reasonable choice.) Moreover, larger $\lambda$ cause more bias to the estimator. In practice, an ordinary least square method or least absolute deviation method could be applied to the selected variables to correct the bias (post $L_1$ PLAD method). We summarized the median of the ordinary LAD estimators on the selected subset in the above table. It can be seen that among the four penalty levels, $\lambda_1$ has the best results in terms of the estimation error $\|\hat \beta-\beta\|_2^2$, and $\lambda_3$ has the best results in terms of post $L_1$ PLAD estimation error. The post $L_1$ PLAD results are very good for all three noise distributions even though the $t(2)$ distribution does not have bounded variance and Cauchy distribution does not have bounded expectation.

\subsection{Gaussian Noise}
Now consider the Gaussian noise case, i.e. $z_i$ are independent and identically normal random variables. The standard deviation $\sigma$ of $z_i$  is varied between 0 and 3. Here we also include the noiseless, where the traditional lasso cannot select the model correctly.  We will use penalty level $\lambda=\sqrt{2n\log p}$ and run 200 times for each value of $\sigma$. For each simulation, we use both the $L_1$ PLAD method and the classical lasso method. For the lasso method, we use $\sigma\times \lambda$ as the penalty level, where we assume the standard deviation is known. In the noiseless case, we use $0.01\times \lambda$ as the penalty level for the lasso method. Here we summaries the average estimation error and the variable selection results of both methods for five different $\sigma$.

\begin{table}[b!]
\centering
\caption{The average of estimation error $\|\hat \beta-\beta\|_2^2$ over 200 replications and the variable selection results for lasso and $L_1$ penalized LAD method.}
\label{Gaussian.noise.table}
\begin{center}
\begin{tabular*}{\textwidth}{@{\extracolsep\fill}|c|c|c|c|c|c|}
\hline Value of $\sigma$ & $\sigma=0$ & $\sigma=0.25$ & $\sigma=0.5$  & $\sigma=1$  &  $\sigma=3$ \\
\hline $L_1$ PLAD: Average of  $\|\hat \beta-\beta\|_2$ & 0 & 0.065  & 0.269  & 1.057 & 8.988 \\
\hline $L_1$ PLAD: Average type I error  & 0 & 0 & 0 & 0 & 0 \\
\hline $L_1$ PLAD: Average type II error  & 0 & 0.185 & 0.150 & 0.120 & 0.175 \\
\hline Lasso: Average of  $\|\hat \beta-\beta\|_2$ & 11.419 & 0.062 & 0.106 & 0.344 & 3.498 \\
\hline Lasso: Average type I error  & 0 & 0 & 0 & 0& 0 \\
\hline Lasso: Average type II error & 24.125 & 0.825 & 0.875 & 0.710 & 0.95 \\
\hline
\end{tabular*}
\end{center}
\end{table}

In table \ref{Gaussian.noise.table}, the average type I error means the average number of significant variables that are unselected over 200 runs. The average type II error means the average number of insignificant variables that are selected over 200 runs. The results show that in terms of estimation, the classical lasso method does better than $L_1$ PLAD method, except the noiseless case. This is partly because that lasso knows the standard deviation and $L_1$ PLAD does not. Also, the penalty level for $L_1$ PLAD method has stronger shrinkage effect and hence cause more bias.

In term of variable selection, the $L_1$ PLAD method does better than classical lasso method. The two methods both select all the significant variables in all the 200 simulations. The $L_1$ PLAD method has smaller average type II errors which means the lasso method tends to select more incorrect variables than the $L_1$ PLAD method. It is worth noting that $L_1$ PLAD method does a perfect job in noiseless case, it selects the perfect model in every run. While the lasso method never have a correct variable selection result.

\subsection{Heavy tail and Heteroscedastic Noise}
In the proof of Theorem \ref{main} and all the discussions, the identically distribution assumption is not essential for our arguments. Now we will study the performance of the $L_1$ PLAD estimator when the noises  $z_i$ are just independent and not identically distributed. We will consider three cases: (a) $z_i\sim N(0, \sigma_i^2)$, where $\sigma_i \sim U(0,3)$ and are independent. (b) $z_i/s_i\sim t(2)$, where $s_i \sim U(0,3)$ and are independent. (c) With probability $1/3$ $z_i\sim N(0, \sigma_i^2)$ and $\sigma_i \sim U(0,3)$, with probability $1/3$ $z_i/s_i\sim t(2)$ and $s_i \sim U(0,3)$, and with probability $1/3$ $z_i/s_i$ follows exponential distribution with parameter 1 and $s_i \sim U(0,3)$ (relocated such that the median is 0). We use penalty $\lambda=\sqrt{2n\log p}$ for all cases. It is worth noting that in all the cases, traditional lasso method and the constrained minimization methods cannot be properly applied since the variances of the noises are unbounded.

\begin{table}[b!]
\centering
\caption{The average of estimation error $\|\hat \beta-\beta\|_2$ over 200 replications and the variable selection results for the $L_1$ PLAD method. Numbers in the parentheses are the medians of the estimation errors of post $L_1$ PLAD method.}
\label{other.noise}
\begin{center}
\begin{tabular*}{\textwidth}{@{\extracolsep\fill}|c|c|c|c|}
\hline  & Case (a) & Case (b) & Case (c)     \\
\hline Average of  $\|\hat \beta-\beta\|_2^2$ & 2.141 (0.253) & 4.355 (0.269) & 2.108 (0.218)  \\
\hline Average type I error  & 0 & 0 & 0.005  \\
\hline  Average type II error  &  0.145 & 0.155 & 0.16 \\
\hline
\end{tabular*}
\end{center}
\end{table}

Table \ref{other.noise} summaries the average estimation errors and variable selection properties of the $L_1$ PLAD method over 200 runs. We also summarize the estimation errors of the post $L_1$ PLAD method in the parentheses. It can be seen that the $L_1$ PLAD method has very nice estimation and variable selection properties for all cases. Compare the variable selection results here with the Gaussian noise case in table \ref{Gaussian.noise.table}, we can see that although we have many different noise structures, the $L_1$ PLAD method can always select a good model. Its variable selection results here are comparable to the Gaussian noise case.

\section{Proofs}
We will first show some technical lemmas and then prove the main results.
\subsection{Technical Lemmas}
We first state the Slastnikov-Rubin-Sethuraman Moderate Deviation Theorem.
Let $X_{ni}, i=1,\ldots,k_n; n \geq 1$ be a double sequence of row-wise independent random variables with $E(X_{ni}) =0$, $E(X^2_{ni})< \infty$, $i=1,\ldots,k_n$; $n\geq 1$, and $B_n^2 = \sum_{i=1}^{k_n} E (X_{ni}^2) \to \infty$ as $n \to \infty$. Let
$ F_n(x) = P \left( \sum_{i=1}^{k_n} X_{ni}< x B_n \right)$. We have
\begin{lemma}(Slastnikov, Theorem 1.1)  If for sufficiently large $n$ and some positive constant $c$,
$$\sum_{i=1}^{k_n} E(|X_{ni}|^{2+c^2}) \rho(|X_{ni}|) \log^{ -(1+c^2)/2}(3+|X_{ni}|) \leq g(B_n) B_n^2,$$
where $\rho(t)$ is slowly varying function monotonically growing to infinity and $g(t) = o(\rho(t))$ as $t \to \infty$, then
$$1 - F_n(x) \sim 1- \Phi(x), F_n(-x) \sim \Phi(-x), \ \ \   n \to \infty,$$
uniformly in  the region $0 \leq x \leq c \sqrt{\log B_n^2}.$
\end{lemma}

\begin{corollary}
\label{MDT.corr}
(Slastnikov, Rubin-Sethuraman) If $q > c^2 + 2$ and
$$\sum_{i=1}^{k_n} E[|X_{ni}|^{q}] \leq K B_n^2,$$
then there is a sequence $\gamma_n\to 1$, such that
$$ \left |\frac{1 - F_n(x) + F_n(-x) }{2(1-\Phi(x))} - 1 \right |  \leq \gamma_n-1 \to 0,  \ \   n \to \infty,$$
uniformly in  the region $0 \leq x \leq c \sqrt{\log B_n^2}.$
\end{corollary}

Remark. Rubin-Sethuraman derived the corollary for $x= t\sqrt{\log B_n^2}$ for fixed $t$. Slastnikov's result adds uniformity
and relaxes the moment assumption. We refer to \cite{S1979} for proofs.

Next, we will state a couple of simple yet useful results. Suppose $U>0$ is a fixed constant. For any $x=(x_1,x_2,\cdots,x_n)\in R^n$, let
$$G(x)=\sum_{i=1}^{n}|x_i|(|x_i|\wedge U),$$
where $a\wedge b$ denotes the minimum of $a$ and $b$. Then we have the following results.
\begin{lemma}
\label{med.1}
For any $x=(x_1,x_2,\cdots,x_n)\in R^n$, we have that
\begin{displaymath}
G(x)\geq \left\{
\begin{array}{cc}
\frac{U\|x\|_1}{2} & \text{ if }\quad \|x\|_1\geq nU/2\\
\|x\|_2^2 & \text{ if }\quad \|x\|_1<nU/2.
\end{array}
\right.
\end{displaymath}
\end{lemma}
\begin{proof}
Let $y=x/U$, then it is easy to see that
\begin{eqnarray*}
\frac{G(x)}{U^2}=\sum_{i=1}^{n}|y_i|(|y_i|\wedge 1).
\end{eqnarray*}
We first consider the case where $\|y\|_1\geq n/2$. Now suppose $|y_i|<1$ for $i=1,2,\cdots,k$ (note that $k$ might be 0 or $n$), and $|y_i|>1$ for $i>k$. Then
\begin{eqnarray*}
\frac{G(x)}{U^2}=\|y\|_1+\sum_{i=1}^{k}y_i^2-\sum_{i=1}^{k}|y_i|\geq \|y\|_1-\frac{k}{4}\geq \frac{\|y\|_1}{2}.
\end{eqnarray*}
Now let us consider the case where $\|y\|_1< n/2$. Suppose there exists an $i$ such that $|y_i|>1$, then there must be a $j$ such that $|y_j|<1/2$. If we replace $y_i$ and $y_j$ by $y_i'=|y_i|-\epsilon\geq 1$ and $y_j'=|y_j|+\epsilon\leq 1/2$ for some $\epsilon>0$, the value of $G(x)/U^2$ decreases. This means that if $G(x)/U^2$ is minimized, all the $y_i$ must satisfy that $|y_i|\leq 1$. In this case, $$G(x)/U^2=\|y\|_2^2.$$
Putting the above inequalities together, the lemma is proved.
\end{proof}

The following lemma is from \cite{CWX2010b}.
\begin{lemma}
\label{norm.ieq} For any $x \in R^n$,
\[
\|x\|_2 - \frac{\|x\|_1}{\sqrt{n}}\le\frac{\sqrt{n}}{4}\big(\max_{1\le i\le n} |x_i|-\min_{1\le
i\le n} |x_i|\big).
\]
\end{lemma}

\begin{remark}
A interesting consequence of the above lemma is: for any $x \in R^n$,
\[
\|x\|_2 \le \frac{\|x\|_1}{\sqrt{n}} + \frac{\sqrt{n}\|x\|_{\infty}}{4}
\]
\end{remark}

\subsection{Proof of Lemma \ref{penalty.cho}}

In this section, we will prove lemma \ref{penalty.cho} by union bound and Hoeffding's inequality. Firstly, by the union bound, it can be seen that
\begin{eqnarray*}
P(c\sqrt{2A(\alpha)n\log p}\leq c\|S\|_{\infty})\leq \sum_{i=1}^{p}P(\sqrt{2A(\alpha)n\log p}\leq |X_i'I|).
\end{eqnarray*}
For each $i$, by Hoeffiding inequality,
\begin{eqnarray*}
P(\sqrt{2A(\alpha)n\log p}\leq |X_i'I|)\leq 2\exp\{-\frac{4A(\alpha)n\log p}{4\|X_i\|_2^2}\}=2p^{-A(\alpha)},
\end{eqnarray*}
since $\|X_i\|_2^2=n$ for all $i$. Therefore,
$$P(c\sqrt{2A(\alpha)n\log p}\leq c\|S\|_{\infty})\leq p 2p^{-A(\alpha)}\leq \alpha.$$
Hence the lemma is proved.

\subsection{Proof of Lemma \ref{A.choice3}}

By the union bound, it can be seen that
\begin{eqnarray*}
P(c\sqrt{n}\Phi^{-1}(1-\alpha/(2p))\leq c\|S\|_{\infty})\leq \sum_{i=1}^{p}P(\sqrt{n}\Phi^{-1}(1-\alpha/(2p))\leq |X_i'I|).
\end{eqnarray*}
For each $i$, from Corollary \ref{MDT.corr},
\begin{eqnarray*}
&&P(\sqrt{n}\Phi^{-1}(1-\alpha/(2p))\leq |X_i'I|)\\
&\leq& 2(1-\Phi(\Phi^{-1}(1-\alpha/(2p))))(1+\omega_n)=\alpha/p(1+\omega_n),
\end{eqnarray*}
where $\omega_n$ goes to 0 as $n$ goes to infinity, provided that $\Phi^{-1}(1-\alpha/2p)\leq (q-2)\sqrt{\log n}$. Hence
\begin{eqnarray*}
P(c\sqrt{n}\Phi^{-1}(1-\alpha/(2p))\leq c\|S\|_{\infty})\leq \alpha(1+\omega_n).
\end{eqnarray*}

\subsection{Proof of Lemma \ref{mean.l1.useful}}

It is easy to see that when $c\geq \frac{6}{a}$,
$$c-\frac{2}{a}\log (1+\frac{a}{2}c)\geq c-\frac{2}{a}\frac{ac}{4}=\frac{c}{2},$$
and when $c\leq \frac{6}{a}$,
$$c-\frac{2}{a}\log (1+\frac{a}{2}c)\geq c-\frac{2}{a}(\frac{ac}{2}-\frac{1}{8}(\frac{ac}{2})^2)=\frac{ac^2}{16}.$$
Similarly, we can show that for any real number $c$,  when $|c|\geq \frac{6}{a}$,
$$E(|z_i+c|-|z_i|)\geq \frac{|c|}{2},$$
and when $|c|\leq \frac{6}{a}$,
$$E(|z_i+c|-|z_i|)\geq \frac{ac^2}{16}.$$
Putting the above inequalities together, the lemma is proved.

\subsection{Proof of Lemma \ref{Uniform.lemma}}

First, it can be seen that for any $1\leq i\leq n$, $||(Xd)_i-z_i|-|z_i||\leq |(Xd)_i|$. So $|(Xd)_i-z_i|-|z_i|$ is a bounded random variable for any fixed $d$. Hence for any fixed $k$ sparse signal $d\in R^p$, by Hoeffding's inequality, we have
$$P\left(B(d)\geq t\right)\leq 2\exp\{-\frac{t^2n}{2\|Xd\|_2^2} \},$$
for all $t>0$. From the definition of $\lambda_k^u$, we know that
$$P\left(B(d)\geq t\right)\leq 2\exp\{-\frac{t^2}{2\lambda_k^u \|d\|_2^2} \}.$$
In the above inequality, let $t=C\sqrt{2k\log p}\|d\|_2$, we have
\begin{equation}
P\left(B(d)\geq C\sqrt{2k\log p}\|d\|_2\right)\leq 2p^{-kC^2/\lambda_k^u},
\end{equation}
for all $C>0$. Next we will find an upper bound for $\sup_{d\in R^P, \|d\|_0=k, \|d\|_2=1}|B(d)|$. We shall use the $\epsilon$-Net and covering number argument. Consider the $\epsilon$-Net of the set $\{d\in R^P, \|d\|_0=k, \|d\|_2=1\}$. From the standard results of covering number, see for example \cite{BM1987}, we know that the covering number of $\{d\in R^k, \|d\|_2=1\}$ by $\epsilon$ balls (i.e. $\{y\in R^k: \|y-x\|_2\leq \epsilon\}$) is at most $ (3/\epsilon)^k$ for $\epsilon<1$. So the covering number of $\{d\in R^P, \|d\|_0=k, \|d\|_2=1\}$ by $\epsilon$ balls is at most $ (3p/\epsilon)^k$ for $\epsilon<1$. Suppose $N$ is such a $\epsilon$-Net of $\{d\in R^P, \|d\|_0=k, \|d\|_2=1\}$. By union bound,
\begin{eqnarray*}
P(\sup_{d\in N}|B(d)|\geq C\sqrt{2k\log p})\leq 2(3/\epsilon)^k p^k p^{-kC^2/\lambda_k^u},
\end{eqnarray*}
for all $C>0$. Moreover, it can be seen that,
\begin{eqnarray*}
\sup_{d_1, d_2\in R^p, \|d_1-d_2\|_0\leq k, \|d_1-d_2\|_2\leq \epsilon} |B(d_1)-B(d_2)|\leq \frac{2}{\sqrt{n}}\|X(d_1-d_2)\|_1\leq 2\sqrt{n}\kappa^{u}_{k}\epsilon.
\end{eqnarray*}
Therefore
\begin{eqnarray*}
&&\sup_{d\in R^P, \|d\|_0=k, \|d\|_2=1}|B(d)|\leq \sup_{d\in N}|B(d)|+2\sqrt{n}\kappa^{u}_{k}\epsilon.
\end{eqnarray*}
Let $\epsilon=\sqrt{\frac{2k\log p}{n}}\frac{1}{2\kappa_k^u}$, we know that
\begin{eqnarray*}
&&P\left(\sup_{d\in R^P, \|d\|_0=k, \|d\|_2=1}|B(d)|\geq C\sqrt{2k\log p} \right)\\
&\leq& P\left(\sup_{d\in N}|B(d)|\geq (C-1)\sqrt{2k\log p} \right)\leq 2(\frac{3p\sqrt{n}\kappa_k^u}{p^{(C-1)^2/\lambda_k^u}})^k.
\end{eqnarray*}
Under the assumption that $p>n$ and $p>3\kappa_k^u$, let $C=1+2C_1\sqrt{\lambda_k^u}$ for some $C_1>1$, we know that
\begin{equation}
P\left(\sup_{d\in R^P, \|d\|_0=k, \|d\|_2=1}|B(d)|\geq (1+2C_1\sqrt{\lambda_k^u})\sqrt{2k\log p} \right)\leq 2p^{-4k(C_1^2-1)}.
\end{equation}
Hence the lemma is proved.

\subsection{Proof of Lemma \ref{l1.deriv}}

Since $||z_i+x|-|z_i||\leq |x|$ is bounded, the expectation always exists. Suppose the density function of $z_i$ is $f(z)$ and $x>0$. It is easy to see that
\begin{eqnarray*}
E(|z_i+x|-|z_i|)&=&\int_{0}^{\infty}f(t)xdt+\int_{-x}^{0}f(t)(2t+x)dt-\int_{-\infty}^{-x}f(t)xdt\\
&=&x(\int_{-x}^{\infty}f(t)dt-\int_{-\infty}^{-x}f(t)dt)+2\int_{-x}^{0}2tf(t)dt\\
&=&x(1-2P(z_i\leq -x))+2\int_{-x}^{0}2tf(t)dt.
\end{eqnarray*}
Hence it is easy to see that $$\frac{d E(|z_i+x|-|z_i|)}{dx}=1-2P(z_i\leq -x).$$

\subsection{Proof of Theorem \ref{main} and \ref{thm.noiseless}}

Now we will bound the estimation error of the $L_1$ penalized LAD estimator. Recall that $h=\beta-\hat\beta$ and $h\in \Delta_{\bar C}=\{\delta\in R^p: \|\delta_T\|_1\geq \bar C\|\delta_{T^c}\|_1\}$. Without loss of generality, assume $|h_1|\geq |h_2|\geq \cdots, \geq |h_p|$. Let $S_0=\{1,2,\cdots,k\}$, we have $h_{S_0}\geq \bar C h_{S_0^c}$. Partition $\{1,2,\cdots, p\}$ into the following sets:
\[
S_0=\{1,2,\cdots, k\}, S_1=\{k+1, \cdots, 2k\},
S_2=\{2k+1, \cdots, 3k\}, \cdots.
\]
Then it follows from lemma \ref{norm.ieq} that
\begin{eqnarray}
\sum_{i\geq 1}\|h_{S_i}\|_2&\leq&
\sum_{i\ge 1}\frac{\|h_{S_i}\|_1}{\sqrt{k}}+\frac{\sqrt{k}}{4}|h_{k+1}|\leq \frac{1}{\sqrt{k}}\|h_{S_0^c}\|_1+\frac{1}{4\sqrt{k}}\|h_{S_0}\|_1\nonumber\\
&\leq&(\frac{1}{\sqrt{k}\bar C}+\frac{1}{4\sqrt{k}})\|h_{S_0}\|_1\leq (\frac{1}{4}+\frac{1}{\bar C})\|h_{S_0}\|_2 \label{basic.2}.
\end{eqnarray}
It is easy to see that
\begin{eqnarray}
&&\frac{1}{\sqrt{n}}(\|Xh+z\|_1-\|z\|_1)\geq \frac{1}{\sqrt{n}}(\|Xh_{S_0}+z\|_1-\|z\|_1)\nonumber\\
&&+\sum_{i\geq 1}\frac{1}{\sqrt{n}}(\|X(\sum_{j=0}^{i}h_{S_j})+z\|_1-\|X(\sum_{j=0}^{i-1}h_{S_{j}})+z\|_1)
\end{eqnarray}
Now for any fixed vector $d$, let
$$M(d)=\frac{1}{\sqrt{n}}E(\|Xd+z\|_1-\|z\|_1).$$

By lemma \ref{Uniform.lemma}, we know that with probability at least $1-2p^{-4k(C_2^2-1)}$,
$$\frac{1}{\sqrt{n}}(\|Xh_{S_0}+z\|_1-\|z\|_1)\geq M(h_{S_0})-C_1\sqrt{2k\log p}\|h_{S_0}\|_2,$$
and for $i\geq1$ with probability at least $1-2p^{-4k(C_2^2-1)}$,
\begin{eqnarray*}
&&\frac{1}{\sqrt{n}}(\|X(\sum_{j=0}^{i}h_{S_j})+z\|_1-\|X(\sum_{j=0}^{i-1}h_{S_j})+z\|_1)\geq M(h_{S_i})-C_1\sqrt{2k\log p}\|h_{S_i}\|_2,
\end{eqnarray*}
where $C_1=1+2C_2\sqrt{\lambda^u_{k}}$ and $C_2>1$ is a constant. Put the above inequalities together, we know that with probability at least $1-2p^{-4k(C_2^2-1)+1}$,
\begin{equation}
\frac{1}{\sqrt{n}}(\|Xh+z\|_1-\|z\|_1)\geq M(h)-C_1\sqrt{2k\log p}\sum_{i\geq 0}\|h_{S_i}\|_2.
\end{equation}
By this and inequality (\ref{basic.1}) and (\ref{basic.2}), we have that with probability at least $1-2p^{-4k(C_2^2-1)+1}$,
\begin{equation}
\label{e1}
M(h)\leq \frac{\lambda\sqrt{k}}{\sqrt{n}} \|h_{S_0}\|_2+C_1\sqrt{2k\log p}(1.25+\frac{1}{\bar C})\|h_{S_0}\|_2.
\end{equation}

Next, we consider two cases. First, if $\|Xh\|_1\geq 3n/a$, then from lemma \ref{med.1} and inequality (\ref{Eofl1}),
\begin{equation}
\label{e2}
\frac{1}{\sqrt{n}}E(\|Xh+z\|_1-\|z\|_1)\geq \frac{3}{16\sqrt{n}}\|Xh\|_1\geq \frac{3\sqrt{n}}{16}\kappa_{k}^{l}\|h_{S_0}\|_2.
\end{equation}
From assumption (\ref{Condition.I}), we must have $\|h_{S_0}\|_2=0$ and hence $\hat\beta=\beta$.

On the other hand, if $\|Xh\|_1<3n/a$, from lemma \ref{med.1} and inequality (\ref{Eofl1}),
\begin{equation}
\label{e3}
\frac{1}{\sqrt{n}}E(\|Xh+z\|_1-\|z\|_1)\geq \frac{a}{16\sqrt{n}}\|Xh\|_2^2\geq \frac{a\sqrt{n}}{16}\eta_{k}^{l}\|h_{S_0}\|^2_2.
\end{equation}
Hence by (\ref{e1}), we know that with probability at least $1-2p^{-4k(C_2^2-1)+1}$,
\begin{equation}
\|h_{S_0}\|_2\leq \frac{16\lambda\sqrt{k}}{na\eta_{k}^{l}}+\sqrt{\frac{2k\log p}{n}}\frac{16C_1(1.25+1/\bar C)}{a\eta_{k}^{l}}.
\end{equation}
In particular, when $\lambda=2c\sqrt{n\log p}$. Putting the above discussion together, we have
\begin{equation}
\|h_{S_0}\|_2\leq \sqrt{\frac{2k\log p}{n}}\frac{16(c\sqrt{2}+1.25 C_1+C_1/\bar C)}{a\eta_{k}^{l}}.
\end{equation}
Since
\begin{eqnarray*}
\sum_{i\geq 1}\|h_{S_i}\|_2^2\leq |h_{k+1}|\sum_{i\geq 1}\|h_{S_i}\|_1\leq \frac{1}{\bar C}\|h_{S_0}\|_2^2,
\end{eqnarray*}
We know that with probability at least $1-2p^{-4k(C_2^2-1)+1}$,
\begin{equation*}
\|\hat\beta-\beta\|_2\leq  \sqrt{\frac{2k\log p}{n}}\frac{16(c\sqrt{2}+1.25 C_1+C_1/\bar C)}{a\eta_{k}^{l}}\sqrt{1+\frac{1}{\bar C}}.
\end{equation*}
where $C_1=1+2C_2\sqrt{\lambda^u_{k}}$ and $C_2>1$ is a constant.

The proof of Theorem \ref{thm.noiseless} is simple. In the noiseless case, we know that
$$\|Xh\|_1\leq \lambda (\|h_{T}\|_1-\|h_{T^c}\|_1) .$$
This means $\|h_{T}\|_1\geq\|h_{T^c}\|_1$ and hence $h\in \Delta_1$. So
$$\|Xh\|_1\geq n\kappa_k^l(1) \|h_T\|_1.$$
Since we assume that $n\kappa_k^l(1)>\lambda$, we must have $\|h\|_1=0$. Therefore $\hat\beta=\beta$.

\bibliographystyle{plain}

\end{document}